\documentstyle[psfig,conf_iap,]{article}

\begin{document}
\heading{%
FUSE Observations towards HD 34078:\\%
Detection of highly excited H$_{2}$ and HD.}
\author{%
Franck Le Petit$^{1}$, Patrick Boiss\'e$^{2}$, Guillaume Pineau des For\^ets$^{1}$, 
Evelyne Roueff$^{1}$, C\'ecile Gry$^{3}$, B-G Andersson$^{4}$,
 Vincent Le Brun$^{3}$.
}
\address{%
Observatoire de Paris-Meudon - DAEC - 5 place Jules Janssen, 92195 Meudon.\\
$ ^{2}$Ecole Normale Sup\'erieure/DEMIRM - France\\%
$ ^{3}$Laboratoire d'astronomie spatiale - France\\%
$ ^{4}$John Hopkins University - USA\\%
}
\begin{abstract}
We present FUSE observations of the extincted O9.5 star, HD 34078. 
The 19 first levels of H$_2$ are detected (i.e. from J=0 to v=1, J=5) 
as well as HD in its two first levels. The excitation of H$_2$ up to 
J=7 can be explained using a combination of Photon Dominated Region
(PDR) and MHD shock models. However, understanding the large amount 
of H$_2$ found in higher excitation states seems to require more
energetic processes that have yet to be identified.     
\end{abstract}

\section{Introduction}
FUSE observations towards HD 34078 have been undertaken to study 
the small scale structure in the spatial distribution of H$_2$ 
(see the paper by Boiss\'e et al. in this volume). 
The first FUSE spectrum turned out to
be exceptionally rich since absorptions from the 19 first levels
of H$_2$ have been identified. After a brief presentation of the line
of sight, we describe the results obtained for H$_2$ and HD. A
preliminary interpretation of the observed excitation is then 
performed using Photon Dominated Region (PDR) and magnetohydrodynamic 
shock models.   

\section{Description of the line of sight to HD 34078}
HD 34078, also known as AE Aurigae, is a O9.5V spectral type star of visual 
magnitude m$_V$= 5.95, located at 500 pc from the sun. 
It is a runaway star with a transverse velocity of 100 km$\cdot$s$^{-1}$ 
and a radial one of 59 km$\cdot$s$^{-1}$. The color excess 
has been estimated to be E$_{B-V}$= 0.52, corresponding
to A$_V$=1.6 for R$_V$=3.1.
Several observations of HD 34078 have already been performed:
HI and CO column densities were measured by Mc Lachlan and 
Nandy \cite{maclac} from IUE observations while CH, CN and C$_2$ have been 
detected by Federman et al. \cite{fed94}. The density is
estimated to be in the range 200 - 500 cm$^{-3}$. In addition, 
CH$^{+}$ and CH, observed by M. Allen \cite{allen}, give a 
Doppler parameter, b $\simeq$ 3 km$\cdot$s$^{-1}$. 
  
\section{Observations}
\begin{figure}
\centerline{\vbox{
\psfig{figure=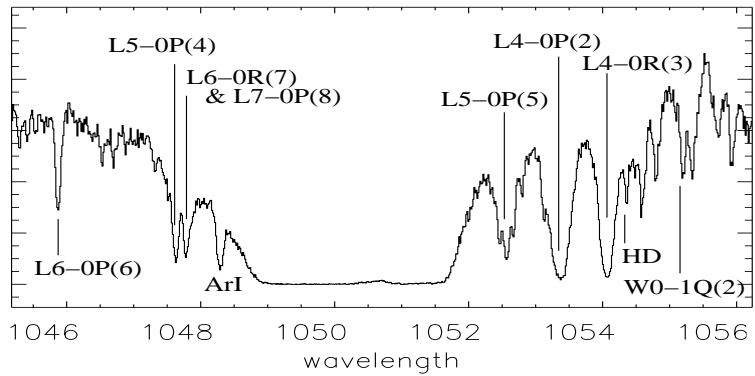,height=5.cm,width=10cm,angle=-90}
}}
\caption[]{Zoom on a 10 m\AA $\,$ interval. }
\label{ident}
\end{figure}

The first observation has been performed in january 2000, with 
a total integration time of 7145s. Data have been processed
and calibrated in a standard way using the FUSE pipeline software.
In the best portions of the LiF1A spectrum, the S/N ratio 
is 30 per 15 m\AA $\,$ pixel; the spectral 
resolution is about 17000. Fig. \ref{ident} displays an extract of the
LiF1A spectrum over a 10 \AA $\,$ interval around 1050 \AA $\,$, 
showing the numerous absorption lines 
present in the spectrum. 
Column densities have been obtained using both a curve of growth
analysis and the fitting program "Owens" written by M. Lemoine. 
Since high resolution CH observations performed in the visible
have revealed no marked velocity structure 
(Federman, private communication),  we assumed that there is only 
one component along the line of sight. 
Most of the observed lines fall on the flat part of the curve of
growth and then poorly constrain the b Doppler parameter. We have thus
adopted the value derived by Allen \cite{allen} for the lowest 
levels of H$_2$ (up to J=4): b=3 km$\cdot$s$^{-1}$.
However, using ten transitions arising from the J=9 state, we have been able 
to build a curve of growth for this level, from which we get 
an upper limit on b(J=9) of 6 km$\cdot$s$^{-1}$. For H$_2$ transitions 
from J=4 to J=9, we have increased the b parameter from 3 to 
5 km$\cdot$s$^{-1}$; an uncertainty of 1 km$\cdot$ s$^{-1}$ on the b
parameter has been considered. 
Error bars are given at three sigma, meaning that values outside of the 
error bars can be excluded. Transitions arising from levels higher
that J=9 are optically thin and thus, corresponding
column densities are independent on any assumption on b; 
the same is true for the J=0 and 1 levels for which absorption lines are damped. 
\section{Results}
We clearly detect absorption lines arising from vibrationally 
excited H$_2$ levels up to v=1 and J=5 (8000 K). Other molecules 
and atoms such as HD, CO, C I, Ar etc... are detected as well.
Absorption lines from levels with still higher excitation energies
(e.g. v=4, J=2) may also be present but, due to the numerous blends 
present, it is difficult to establish unambiguously their
identification. The table displays the column densities derived for the 19 first energy levels sorted by increasing energy. We derive a total molecular 
hydrogen column density of:
\begin{displaymath}
N(H_2)= 6.4 \times 10^{20} cm^{-2}
\end{displaymath}

\begin{table}
\begin{tabular}{c |c c c c || c | c c c}
 Energy (K) & v & j & N (cm$^{-2}$) & &Energy (K) & v & j & N (cm$^{-2}$)\\
 \hline
      0& 0& 0& 3.20$\times 10^{20}$& &6140.2& 1& 1& 2.10$\times 10^{14}$ \\
  170.5& 0& 1& 3.20$\times 10^{20}$& &6471.6& 1& 2& 8.30$\times 10^{13}$ \\
  509.8& 0& 2& 1.80$\times 10^{19}$& &6951.6& 1& 3& 1.70$\times 10^{14}$ \\
 1015.1& 0& 3& 6.20$\times 10^{18}$& &7197.0& 0& 9& 5.98$\times 10^{14}$ \\
 1681.7& 0& 4& 7.10$\times 10^{17}$& &7584.6& 1& 4& 7.00$\times 10^{13}$ \\
 2503.9& 0& 5& 3.30$\times 10^{17}$& &8365.3& 1& 5& 1.80$\times 10^{14}$ \\
 3474.4& 0& 6& 4.00$\times 10^{15}$& &8677.3& 0&10& 3.99$\times 10^{13}$ \\
 4586.4& 0& 7& 2.50$\times 10^{15}$& &9286.6& 1& 6&    ---               \\
 5829.8& 0& 8& 6.00$\times 10^{14}$& &10261.8& 0&11& 4.00$\times 10^{13}$\\
 5987.1& 1& 0& 2.20$\times 10^{13}$& &       &     &                     
  \label{table1}
\end{tabular}
\end{table}  

Seven lines from HD J=0 are clearly detected but some of them are
blended and it appears difficult to fit all lines with a single N(HD)
value. We get the approximate value, N(HD J=0) $\simeq$ 10$^{15}$
cm$^{-2}$; the 
accuracy is too low to derive any D/H ratio. 
The weak HD L7-0R(1) transition
arising from the J=1 level is tentatively detected at 1022 \AA$\,$ giving 
N(HD J=1)) = 1.0 (-0.4 +0.8) $\times$ 10$^{14}$ cm$^{-2}$. A previous 
detection of excited HD was obtained using Copernicus by Wright and 
Morton \cite{wright79} towards $\zeta$ Oph.
\begin{figure}
\centerline{\vbox{
\psfig{figure=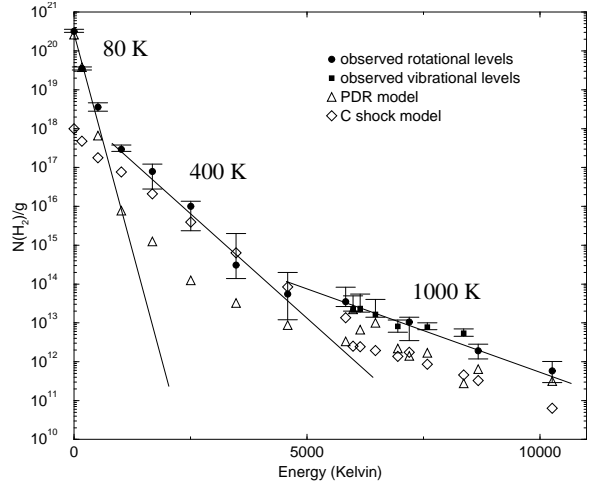,height=6.5cm,angle=-90}
}}
\caption[]{H${_2}$ excitation diagram. For the C shock model
considered, a velocity of 25 km$\cdot$s$^{-1}$ and a magnetic field of
10 $\mu$G have been assumed.}
\label{excitation}
\end{figure}
\section{Discussion}
Figure \ref{excitation} displays the excitation diagram of H$_2$.
The 3 first H$_{2}$ levels are consistent with a temperature of 80
K, and a thermal equilibrium ortho-to-para ratio. For higher excited
levels, the distribution can be artificially decomposed in two thermal
ones with excitation temperatures of 400K (for levels between J=3 and J=7) 
and 1000K (beyond J=7). Using first the PDR model of Le Bourlot et
al. \cite{lebou93} and adopting a density of 500 cm$^{-3}$ (constrained by the 
observations of Federman et al \cite{fed94}) together with a radiation field 
of 2 times the ISRF, it is possible to reproduce the column 
density of the two first H$_2$ levels and other molecules, except
CH$^+$.  However, this model considerably underestimates 
column densities for levels J $\geq$ 3.
A more energetic mechanism is required to account for the latter. 
We have run several C shock models in the presence of a radiation
field (reference \cite{montei}). Reasonable agreement is obtained for levels 
J=3 to J=7 with a shock velocity of 25 km$\cdot$s$^{-1}$ and a magnetic field of 
10 $\mu$G. But again, the large amount of H$_2$ observed  at J $\geq$ 8 is well
above the prediction of these models. 
One plausible scenario could involve H$_2$ located in a bow shock
located around HD 34078. Since it is a runaway star, it is now well
detached from the cloud where it formed, unlike HD 37903 towards which
Meyer et al. \cite{meyer} found a lot of H$_2$ excited to high energies by
fluorescence. However, due to the large velocity of HD 34078, a strong
bow shock is expected to be present at the interface of the stellar 
wind and of the ambient ISM (Mc Low et al. \cite{maclow}). Gas trapped in 
this dense layer might contribute to the observed absorption 
- a scenario that we need to explore in more details.

\end{document}